\documentclass[a4paper]{article}
\pdfoutput=1
\usepackage{subfig}

\usepackage{INTERSPEECH2020}

\title{Relational Teacher Student Learning with Neural Label Embedding for Device Adaptation in Acoustic Scene Classification}
\name{Hu Hu$^{1}$, Sabato Marco Siniscalchi$^{2}$, Yannan Wang$^{3}$, Chin-Hui Lee$^{1}$ }
\address{$^1$Electrical and Computer Engineering, Georgia Institute of Technology, Atlanta, GA, USA \\
$^2$Computer Engineering School, University of Enna, Italy \\
$^3$Tencent Media Lab, Tencent Corporation, Shenzhen, Guangdong, China}
\email{huhu@gatech.edu, marco.sinsalchi@unikore.it, yannanwang@tencent.com, chl@ece.gatech.edu}

\begin{document}

\maketitle
\begin{abstract}
In this paper, we propose a domain adaptation framework to address the device mismatch issue in acoustic scene classification leveraging upon neural label embedding (NLE) and relational teacher student learning (RTSL).  Taking into account the structural relationships between acoustic scene classes, our proposed framework captures such relationships which are intrinsically device-independent. In the training stage, transferable knowledge is condensed in NLE from the source domain. Next in the adaptation stage, a novel RTSL strategy is adopted to learn adapted target models without using paired source-target data often required in conventional teacher student learning. The proposed framework is evaluated on the DCASE 2018 Task1b data set. Experimental results based on AlexNet-L deep classification models confirm the effectiveness of our proposed approach for mismatch situations.
NLE-alone adaptation compares favourably with the conventional device adaptation and teacher student based adaptation techniques. NLE with RTSL further improves the classification accuracy.
\end{abstract}

\section{Introduction}
\label{sec:intro}

In recent years, we have witnessed a great progress in the acoustic scene classification (ASC) task, as demonstrated by the high participation in the IEEE Detection and Classification of Acoustic Scenes and Events (DCASE) challenges \cite{dcase2016, dcase2017, dcase2018}. Top ASC systems use deep neural networks (DNNs), and the main ingredient of their success is the application of deep convolutional neural networks (CNNs) \cite{asc-cnn1, asc-cnn2, asc-cnn3, dcase-2020-rank2, asc-cnn4, asc-asm}. Further boost in ASC performance is obtained with the introduction of advanced deep learning techniques, such as attention mechanism \cite{asc-attention1, asc-attention2, asc-attention3}, mix-up \cite{asc-mixup2, asc-mixup1}, Generative Adversial Network (GAN) and Variational Auto Encoder (VAE) based data augmentation  \cite{asc-gan1, asc-gan2}, and deep feature learning \cite{openl3, asc-xvector, asc-feature1, asc-feature2}. Nevertheless, those ASC systems yet do not work well when processing audios from mismatched domain, e.g., audios recorded with different devices \cite{dcase2018}.  Device mismatch is an inevitable problem in a real production, and it is therefore an important aspect to handle when deploying an ASC system. Indeed, a new sub-task, namely \emph{Task1b}, has been added to DCASE 2018 \cite{dcase2018} to foster research in that  direction. The goal is to design a system that can attain a good performance on 10-second audios segments collected with target devices, which are either not represented at a development phase, or represented during the ASC system deployment with a scarce amount of training material compared to that available for the source device. However, Task1b attracted only a minor interest among DCASE 2018 and 2019 participants, and even fewer teams were directly concerned with the device mismatch issue.

In the literature, there exist a few approaches that tackle the domain invariant problem in ASC. For example, multi-instance learning \cite{asc-mil}, and low-level or mid-level feature learning \cite{asc-ivector, openl3, asc-mixup1}, which however address the robustness issue in a broader sense.  Less approaches have instead been proposed to directly combat the ASC device mismatch issue, which is actually the focus of the present work. In particular, spectrum correction \cite{asc-speccorr} and channel conversion \cite{asc-da-conversion} build a front-end module to convert speech features from the source domain to target domain before feeding them to the back-end classifier. Besides front-end features, mid-level feature based transfer systems, which uses bottleneck features \cite{asc-bn-transfer} or hidden layer representations \cite{asc-da-2019rank4} are adopted to transfer knowledge from source to target domain. Adversarial training methods in \cite{asc-adda1, asc-adda2} leverage an extra domain discriminator to solve the device mismatch problem although the key focus is on lack of labeled target data.

Teacher-student (TS) learning, also named as knowledge distillation \cite{Hinton2015}, has recently been shown to be effective in ASC and other domain adaptation speech tasks, e.g., \cite{asc-ts, ts1, ts-jinyu, ts-zhong}. The key idea is to minimizes the distance measurement between teacher and student model output distributions, i.e., the information is transferred at a soft-label level.
In \cite{rkd},  relational knowledge distillation (RKD) is demonstrated to improve the knowledge distillation process. RKD takes into account the relations of outputs rather than individual outputs themselves.
Unfortunately, conventional TS learning can be applied with success if: (i)  source and target data is from the same or similar domain \cite{asc-ts, ts-gan}, or (ii) source and target data come in pair although belong to different domains \cite{ts-jinyu, ts-zhong, ts-wei}. Neural label embedding (NLE), recently proposed in \cite{le}, is an ingenues solution to distill knowledge across domains when neither of the aforementioned two requirements could be met.
NLE can be viewed as the centroid of soft labels from the same class. As to extension of soft labels, it encodes the knowledge distilled from the source domain and teacher model, which can then be transferred to the target domain. 


In this study, we extend the NLE adaptation scheme \cite{le} by taking into account relationships among different acoustic scenes during adaptation. We achieve this goal by proposing a relational teacher student learning (RTSL) approach based on NLE for ASC device mismatching problem. First, NLE is learned from a relatively large-size source data set, i.e., collected with the source devices.
Next, ASC system is adapted to the target device leveraging upon target domain data only, i.e., teacher-student learning with unpaired data, and the set of NLE, one each per acoustic scene class. The proposed solution is assessed against the DCASE 2018 Task1b data. Experimental results confirm our intuitions and demonstrate that our adaptation technique generates a significant classification improvement on target domain data. Indeed, NLE-based TS adaptation outperforms both (i) multi-device training strategies, and (ii) conventional TS adaptation schemes. Furthermore, an additional boost is obtained when TS adaptation is carried out leveraging structural information.

\begin{figure}[t]
  \centering
  \includegraphics[width=0.65\linewidth,height=4.5cm]{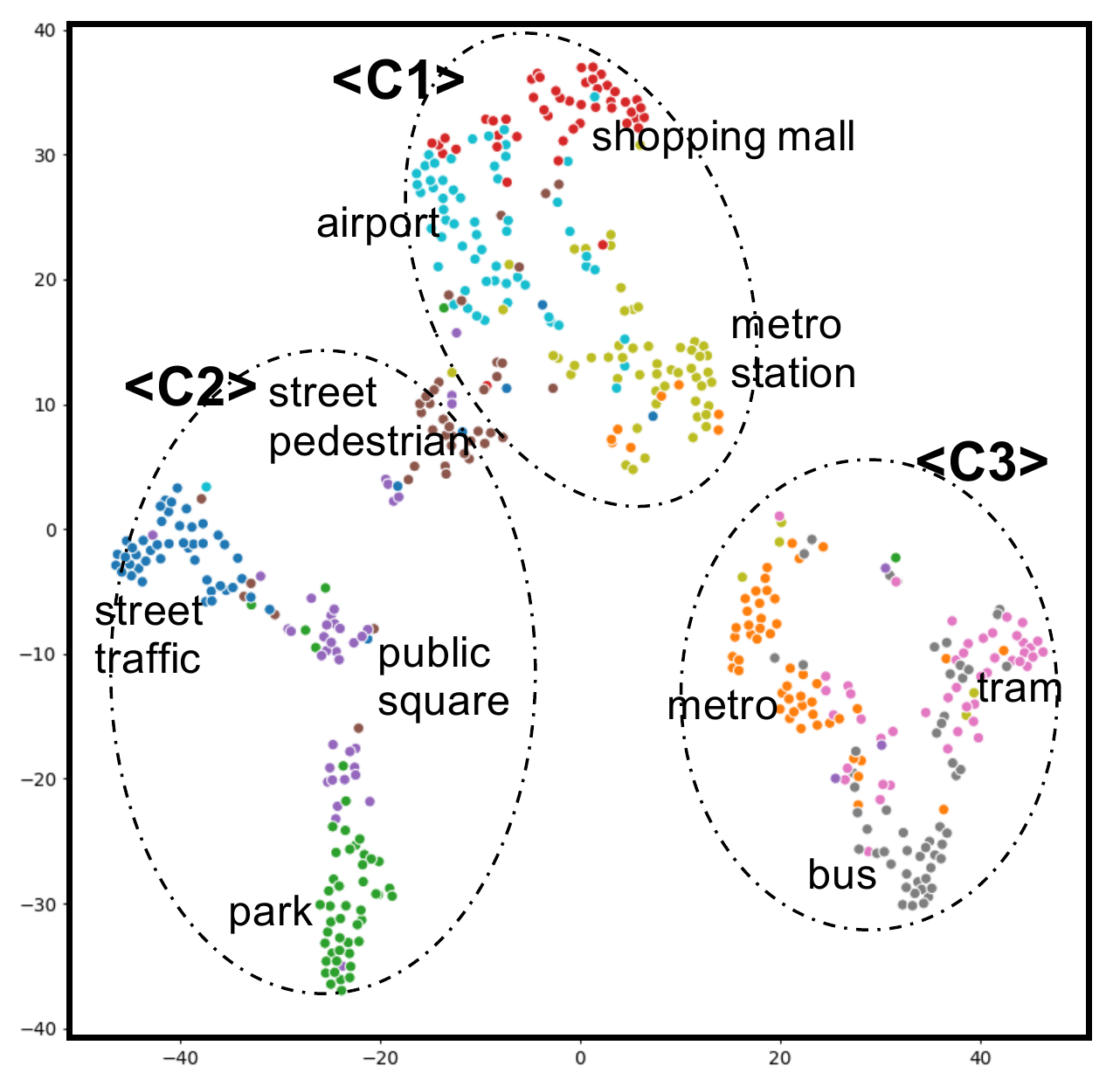}
  \caption{SKLD based t-SNE plot of an AlexNet-L model outputs, i.e., posterior probabilities. $C1-C3$ indicate public in-door area, public out-door area and transportation clusters, respectively.}
  \label{fig:relationships1}
  \vspace{-1mm}
\end{figure}



\section{Neural Label Embedding for Adaptation}
\label{sec:le}
We focus on transfer learning of structural relationships among ASC classes. Conventionally one-hot label is used to train the ASC system, assuming the similarities between each class pair to be the same at the label level. To visualize the distance between two probability densities, we use SKLD based t-SNE \cite{Maaten2008}, an extension of Stochastic neighbor embedding (SNE) \cite{Hinton2003}, to transform the observations in high-dimensional space into a low-dimensional space which preserves neighbour identities by minimizing the symmetrical Kullback-Leibler divergence (SKLD) of the pairwise distributions between the two spaces where Student t distribution is assumed. In Figure~\ref{fig:relationships1}, the points are 2-dim representations of audio samples, recorded with Device A, scattering all over the figure. On the other hand, three clusters, labeled C1, C2 and C3, representing public in-door, public out-door and transportation areas, respectively, can be observed. This confirms our conjecture and intuition that the intrinsic relationships between different acoustic scenes depend more on the recording environments. 

\subsection{Source Model Training}
\label{sec:2a}
\begin{figure}[ht]
  \centering
  \includegraphics[width=0.95\linewidth]{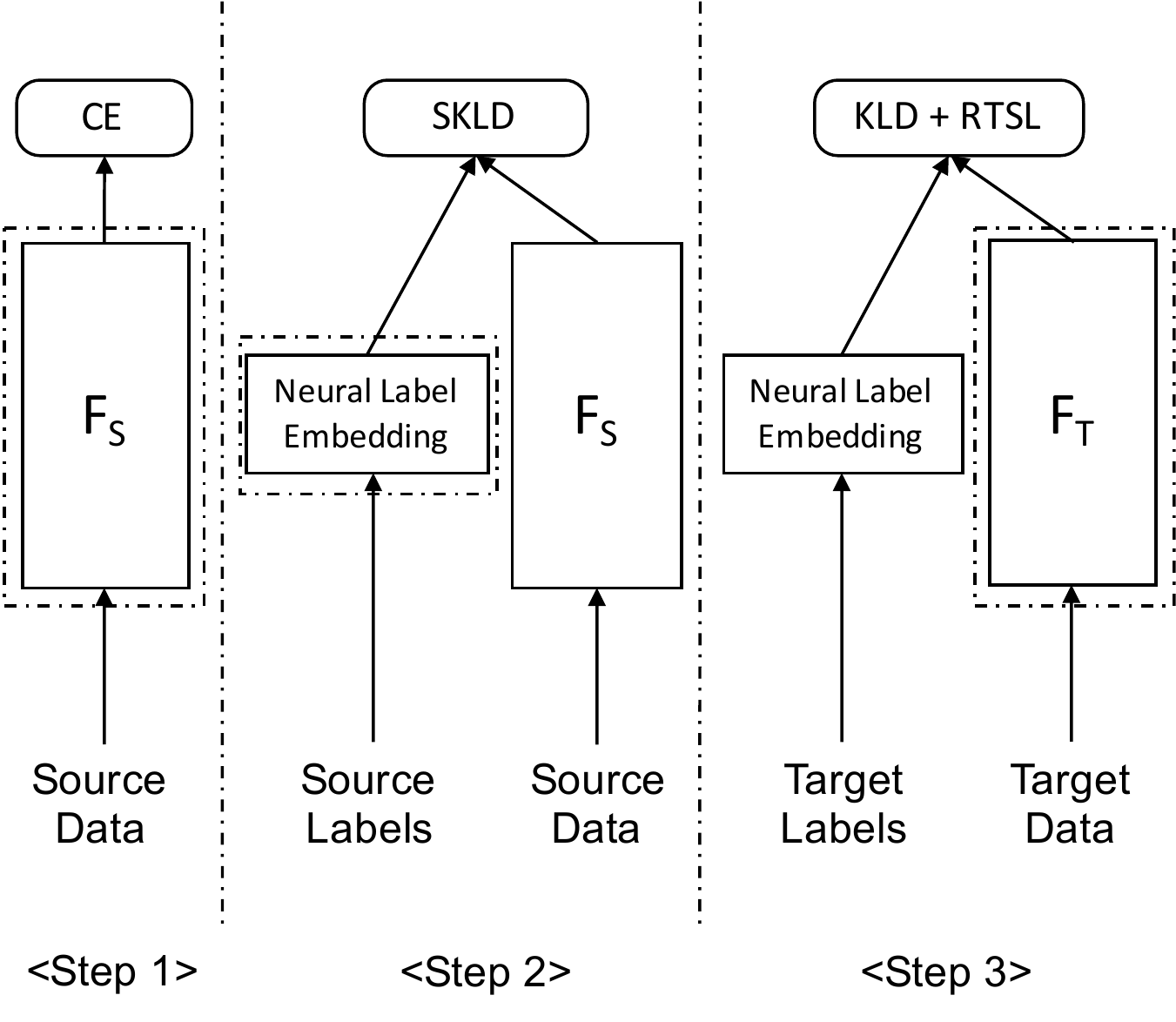}
  \caption{Framework of NLE with RTSL. The parameters of the models in the dashed boxes in Steps 1-3 are learnable.}
  \label{fig:le}
  \vspace{-1mm}
\end{figure}

We now describe NLE's generation and usage as suggested in \cite{le}. NLE starts with building a stand-alone deep model for the ASC task, as shown in the $<$Step 1$>$ of Figure \ref{fig:le}. Here, the source data, $X^S = \{x^S_1, x^S_2, ..., x^S_{N_S}\}$, and the related one-hot class label vectors, $Y^S = \{y^S_1, y^S_2, ..., y^S_{N_S}\}$, are employed to estimate the model parameters with a Cross Entropy (CE) loss:

\begin{equation}
    L_{CE} (X^S, Y^S) = -\frac{1}{N_S} \sum_{x^S_i, y^S_i \in D_S} \sum^K_{j=1} y^S_{i,j}  \log F_S(x^S_i)_j,
    \label{eq:loss_ce}
\end{equation}
where $D_S$ indicates source domain data, $N_S$ is the number of input samples, K is the number of output classes which is also the dimension of the one-hot label vector, and $F_S$ represents the source domain model, performing input-output mapping.

\subsection{NLE Label Generation}
\label{sec:2b}
NLE can function as labels in place of the one-hot class labels for the source data. The goal is to distill the knowledge of $F_S$ into a dictionary of NLE vectors, one for each acoustic scene class, which is predicted at the output layer in $<$Step 2$>$ of Figure~\ref{fig:le}.
Each K-dim NLE vector for class $c$, $NLE_{c}$, encodes the statistical relationship between class $c$ and all other classes via the output distributions of the source-domain $F_S$ given all features aligned with class $c$ at the input. 
Thus, NLE is generated by SKLD ($\gamma_2$), which is to measure the distance between two distributions, $P$ and $Q$, defined in the following:

\begin{equation}
    \gamma_1 (P, Q) = \sum_{p_i, q_i \in P,Q} p_i \log \frac{p_i}{q_i},
\end{equation}
\begin{equation}
    \gamma_2 (P, Q) = \frac{1}{2} (\gamma_1 (P, Q) + \gamma_1  (Q, P)).
\end{equation}
 A $K \times K$ NLE matrix, $[NLE_1, NLE_2, ..., NLE_K]$, denoted as $NLE_M$, can thus be trained using SKLD. The associated loss function $L_{LE}$ is defined as follows:
\begin{equation}
    L_{LE} = \frac{1}{N_S} \sum_{x^S_i, y^S_i \in D_S} \gamma_2 (F_S (x^S_i), NLE_M \cdot y_i^S)
\end{equation}
\noindent where $NLE_M \cdot y_i^S$ represents the label embedding vector of input $x_i^S$ with the one-hot label vector $y_i^S$. The matrix multiplication "$\cdot$" is an indexing operation on $NLE_M$. It should be noted that all elements in the vector $NLE_c$ should sum to 1.

\subsection{NLE for Device Domain Adaptation}
\label{sec:2c}
The final step to accomplish device adaptation, as indicated in $<$Step 3$>$ in Figure~\ref{fig:le}, is to obtain the adapted target model, $F_T$, starting from the seed source model, $F_S$, using only the trained $NLE_M$ and target domain data $D_T$ with available inputs $X^T = \{x^T_1, x^T_2, ..., x^T_{N_T}\}$ and labels $Y^T = \{y^T_1, y^T_2, ..., y^T_{N_T}\}$. This is different from conventional TS, which requires paired source-target data, including soft labels obtained with $F_S$. In our case,  $Y^T$, the one-hot labels available for the target data, are used to fetch and be replaced by the corresponding NLE in $NLE_M$. Finally, KLD can be used to train $F_T$ with the following criterion:
\begin{equation}
    L_{NLE} (X^T, Y^T) = \frac{1}{N_T} \sum_{x^T_i, y^T_i \in D_T} \gamma_1(NLE_M \cdot y_i^T, F_T(x_i^T)).
    \label{eq:loss_le}
\end{equation}

It should be noted that with a relatively large-scale target domain data, the simplest way to adapt $F_S$ to the target condition is to use the target data and one-hot labels $Y^T$ to fine-tune the original model using the CE loss. However, that easily leads to over-fitting as the amount of target data decreases. The model to be used for TS learning in $<$Step 3$>$ will be presented next.


\section{Relational Teacher Student Learning with Neural Label Embedding}
\label{sec:rtsl}
As aforementioned, for conventional TS learning for domain adaptation, soft labels generated from teacher are employed to train the student with the KLD. With well-trained source model $F_S$, the loss $L_{TS}$ for TS learning is
\begin{equation}
    L_{TS} (X^T, Y^T) = \frac{1}{N_T} \sum_{X^{S'}_i \in D_S, X^T_i \in D_T} \gamma_1( F_S(x^{S'}_i), F_T(x^T_i)),
\end{equation}

\noindent where $X^{S'}$, in the case of acoustic scene classification, is often a subset of $X^S$ owing paired target data, $X^T$. In this work, paired data represent two acoustic scenes collected in the same location and time but with two different devices. Traditional TS learning encodes the knowledge from source domain data and source model into soft labels. Benefiting from this, knowledge can be transferred from the source domain to target domain.


Taking into account the structural relationships among output classes has shown to be beneficial in \cite{rkd}. We therefore propose to use relational teacher student learning (RTSL) to carry out device adaptation and generate the target model leveraging upon NLE, and their structural relationships. Using ASC task as an example, as visualized in  Figure~\ref{fig:relationships1}, the measured distance between sounds from metro and park should be larger than the sounds from metro and train. This property is from the feature of acoustic scene itself, and it should not be modified by the nature of the device used during the recording phase. Thus, we can minimize the distance of each class between source and target domain data. In our experiments, the SKLD is used as the measurement of distance among probability distributions.

The total mutual distance value, $V$, of the target model output probabilities, $F_T(X^T)$, is obtained considering all possible $N^2$ pairs in the training set. $V$ is thus givn in Eq. (\ref{mutdist}).
\begin{equation}
    V (F_T(X^T)) = \frac{1}{N_T^2} \sum_{x^T_i \in D_T} \sum_{x^T_j \in D_T} \gamma_2 (x^T_i, X^T_j).
    \label{mutdist}
\end{equation}
During RTSL, we minimize the divergence between total mutual distance of output and NLE. In our experiments, we use smoothed L1 loss, a.k.a. Huber loss \cite{Huber1964}, which is given by
\begin{equation}
SL1 (x, y)=\left\{
\begin{aligned}
\frac{1}{2} (x - y)^2, |x - y| \leq 1, \\
|x - y| - \frac{1}{2}, |x - y| > 1.
\end{aligned}
\right.
\end{equation}
Therefore, the loss for RTSL is
\begin{equation}
    L_{RTSL} = SL1 ( V(NLE_M), V(F_T(X^T))).
    \label{eq:loss_rtsl}
\end{equation}

During target model training of relational teacher student learning with NLE, we use the linear combination of $L_{NLE}$ in Eq.~(\ref{eq:loss_le}) and $L_{RTSL}$ in Eq.~(\ref{eq:loss_rtsl}), which is given by
\begin{equation}
    L_{NLE-RTSL} = L_{NLE} + \lambda * L_{RTSL}
    \label{eq:loss_all}
\end{equation}
where $\lambda$ is a tunable hyper-parameter to balance the loss terms. According to our experiments, this parameter is very robust to the model training. We set $\lambda=10$ to make the two contributors of our loss have the same scale range.

\vspace{0.2cm}

\section{Experiments and Analysis}
\label{sec:exp}
\subsection{Experimental Setup}
The proposed adaptation framework is evaluated on the DCASE 2018 Task1b development data set \cite{dcase2018}. Task1b is provided with 28 hours of acoustic scene audio recordings acquired with three real devices, namely  device A (24 hours), device B (2 hours) and device C (2 hours). Audio segments are recorded in 10 different acoustic scenes, and each audio segment lasts 10 seconds. For each 10-second single-channel audio segment, short-time Fourier transofrm (STFT) with 2048 FFT points is applied, using a window analysis of  25ms and a shift of 10ms. Mel filter-banks with 128 bins are used to extract the 128-dimension log-Mel filter bank  energy (128-D LMFB) features. In our experiments, we follow the officially recommended train-test partition of the data. Training data from device A is regarded as the source domain data, and device B and device C are regarded as target domain. It should be noticed that our goal is slightly different from that for the  Task1b. We regard  data from devices B and C as two separate target domains. We aims to solve the device mismatch issue for one specific target device at a time, which is a more common scenario in real applications.

All ASC deep models used in our experiments use an AlexNet \cite{alexnet} based CNN structure. Resource constraints impose us to reduced the amount of neural parameters, and we refer to this model as AlexNet-L, and it's implemented in PyTorch. AlexNet-L has five convolutional layers with the kernel size of $4 \times 4$ and two fully connected layers with the hidden dimension of $1024$. Each convolutional layer consists of a convolution operation, a batch normalization block, non-linear processing block with ReLU activation function, and a max pooling block. 
Each input audio recording is chopped into segments of  20 frames  (0.2 seconds per segment). Therefore, final scene classification is  obtained through majority voting among all segments. AlexNet-L is trained with stochastic gradient descent (SGD) algorithm with cosine based learning rate scheduler. The initial learning rate is set to 0.01 for baseline training, and 0.002 for adaptations on target device. The temperature parameter \cite{ts1} is used and set to 2.0 in all soft labels and NLE-based experiments. The linear combination parameter $\lambda$ in Eq. (\ref{eq:loss_all}) is set to 10 to balance the loss terms.

\vspace{0.1cm}
\subsection{Experimental Results}
 Table~\ref{tab:res} shows our experimental results on  DCASE 2018 Task1. From the top three rows in Table~\ref{tab:res}, we can observed that device mismatch is indeed an critical aspect in acoustic scene classification. The official baseline system \cite{dcase2018}, and the AlexNet-L (All Devs), listed in the two first rows, are trained using  data from all the three devices, namely A, B, and C. The amount of Device A training material is much larger than that available for Devices B \& C (24 hours against 2 hours); therefore, when the baseline systems are skewed toward Device A, and the  degradation in the classification accuracy when moving from Device A test data to  device B \& C test data is significant. In the third row in Table~\ref{tab:res}, we report acoustic scene classification accuracy when AlexNet-L is trained only on device A data (this is, Device B \& C data is missing). This experiment is conducted to clarify any doubt about the severity of the device mismatch issue. In fact, the accuracy drops from 69.6\% on device A  down to 13.4\% and 16.2\% on Devices B and C, respectively.
\begin{table}[h]
\centering
\caption{Evaluation results on DCASE 2018 Task1b.}
\label{tab:res}
\begin{tabular}{l||c|c|c}
\hline
\hline
Model                     & \begin{tabular}[c]{@{}l@{}}Dev A \\ acc.(\%)\end{tabular} & \begin{tabular}[c]{@{}l@{}}Dev B \\ acc.(\%)\end{tabular} & \begin{tabular}[c]{@{}l@{}}Dev C \\ acc.(\%)\end{tabular} \\
\hline
\hline
Official Baseline \cite{dcase2018}         & 58.9                                                  & 45.6                                                  & 52.3                                                  \\
AlexNet-L (All Devs)      & 67.0                                                  & 52.5                                                  & 54.7                                                  \\

AlexNet-L (Dev A)         & 69.6                                                  & 13.4                                                  & 16.2                                                  \\
\hline
AlexNet-L (Dev A)  & & & \\
\,\,\,+ one-hot adaptation    & -                                                     & 57.0                                                  & 60.8                                                  \\
\,\,\,+ soft labels adaptation & -                                                     & 56.7                                                  & 60.6                                                  \\
\,\,\,+ NLE adaptation          & -                                                     & 58.7                                                  & 62.0                                                  \\
\,\,\,+ NLE-RTSL adaptation    & -                                                     & \textbf{59.2}                                                 & \textbf{64.2}      \\
\hline
\hline
\end{tabular}
\vspace{-1mm}
\end{table}

In Section~\ref{sec:le}, we pointed out that a simple yet viable domain adaptation approach with labeled adaptation data is to fine-tune the source deep  model using adaptation data and  one-hot labels. We put forth this supervised adaptation scheme using AlexNet-L (Dev A), trained on Device A data only, as a seed model, and and Device B \& C data (separately) along with one-hot class labels, and the  CE loss in Equation~\ref{eq:loss_ce}. From Table~\ref{tab:res}, we can observe that  one-hot adaptation improves performance on target devices (see 4th row), and classification accuracy goes from 13.4\% to 57.0\% for Device B  and from 16.2\% to 60.8\% for Device C. By comparing with multi-device training (2nd row), we can conclude that one-hot adaptation outperforms multi-device training.

To better appreciate our contribution, the conventional TS learning, dubbed soft labels adaptation in Table \ref{tab:res}, is also implemented. This adaptation approach can be put forth because Device B \& C data are paired with some of Device A data. The teacher is AlexNet-L (Dev A). The student is obtained by copying the teacher architecture and then adapting it using target device data and soft labels generated by the teacher model fed with the Device A paired data. In soft labels adaptation, the teacher parameters remain unchanged. From the results in the 4rd, and 5th rows in Table \ref{tab:res}, we can observe that slightly worse results are obtained using TS with soft labels adaptation than using a simpler one-hot adaptation solution. That may be due to the the mismatch between source and target domain is large.


Finally, we turn our attention to the key architecture proposed for device adaptation. In the last two rows in Table~\ref{tab:res}, experimental results of TS adaptation with NLE (LE adaptation) and NLE with RTSL framework (NLE-RTSL adaptation) are reported. First, we can observe that both NLE adaptation approaches attain superior performance compared to one-hot adaptation.  For Device B, classification accuracy goes from 57.0\% to 58.7\%, and from 60.8\% to 62.0\% for Device C. A significant boost in the classification results can further be obtained by imposing structural constrains during knowledge distillation. Indeed, NLE-RTSL adaptation allows us to attain top performance of 59.2\% for Device B, and 64.2\% on Device C, respectively.

\begin{figure}[t]
  \centering
  \includegraphics[width=0.65\linewidth,height=4.5cm]{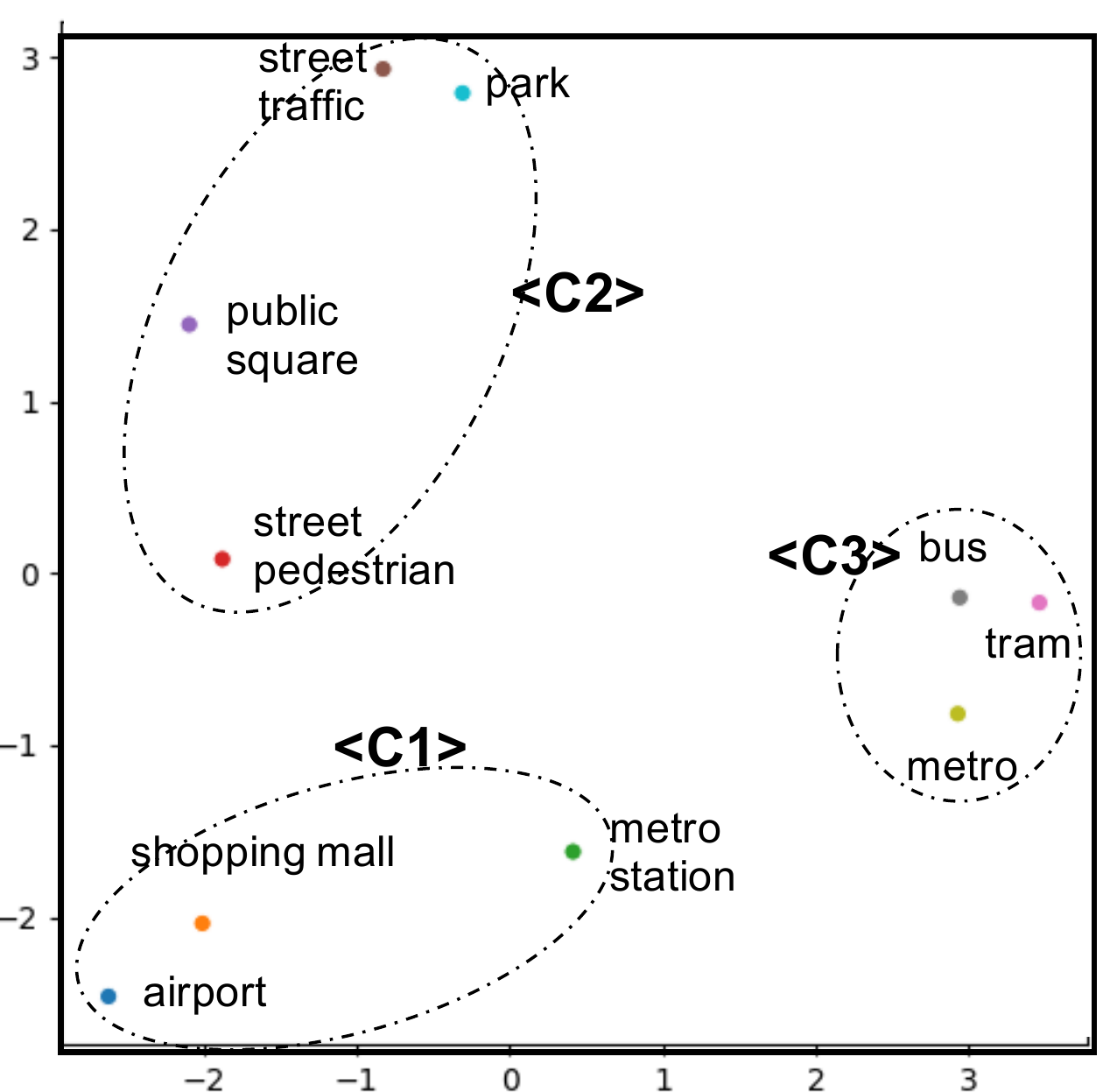}
  \caption{Scatter plot on two-dimensional space of NLE vectors. $C1-C3$ indicates the three clusters of public in-door area, public out-door area and transportation, respectively.}
  \label{fig:relationships2}
  \vspace{-1mm}
\end{figure}

\vspace{0.1cm}

\subsection{Visual Analysis}
A visual analysis can shed light on the nature of our proposed NLE for acoustic scenes. Recall in Figure~\ref{fig:relationships1} we show audio samples belonging to Device A obtained with SKLD based t-SNE method. It is easy to see that three main clusters arise. They are circled with a dashed ellipses labeled as C1, C2 and C3, for public in-door, public out-doors, and transport-related areas, respectively. A close visual inspection immediately reveals some key features of those acoustic scenes. For instance,  metro samples are close to the metro station samples, which are both semantically related to the metro concept. In the same way, street pedestrian samples are close to street traffic samples. In addition, street pedestrian is also close to the public in-door area, since both scenes can capture walking people. Otherwise, street pedestrian scene is far away from bus-, metro-, and tram-related acoustic scenes. However, within classes we can observe some overlap, and that because not all soft labels are equally good. In Figure~\ref{fig:relationships2}, we plot the NLE vectors, one centriod per acoustic scene. NLE are generated from the source Device A training data and Principal component analysis (PCA) is applied for dimension reduction. Comparing Figures~\ref{fig:relationships1} and Figure~\ref{fig:relationships2}, we can see that the three main clusters are present in both plots, yet we have a single NLE per acoustic scene which makes our proposed adaptation approach less sensitive to the quality of the soft labels. In summary, the two plots reveal the correctness of our conjecture for an existence of structural relationships between acoustic scenes. Moreover, NLE encodes these structural relationships in a compact way in low-dimensional spaces.


\section{Conclusion}
\label{sec:con}
In this paper, a relational teacher student learning framework with neural label embedding is proposed to resolve the device mismatch issue in acoustic scene classification. We explore the similarities or dissimilarities between pairs of classes. This structural relationship is learned and encoded into NLE and then transferred from the source device domain to the target device domain via the relational teacher-student approach. Our proposed framework is assessed against the DCASE 2018 Task1b development data set, and experimental results demonstrate not only the viability of our approach, but also that a significant improvement of the classification accuracy on the target device data can be obtained. Furthermore, a visual analysis is provided to shed light on the key characteristics of the proposed neural label embedding concept.
\clearpage
\bibliographystyle{IEEEtran}

\bibliography{mybib}

\end{document}